# Denial of service attack in the Internet: agent-based intrusion detection and reaction

Oleksii Ignatenko

**Abstract**. This paper deals with denial of service attack. Overview of the existing attacks and methods is proposed. Classification scheme is presented for a different denial of service attacks. There is considered agent-based intrusion detection systems architecture. Considered main components and working principles for a systems of such kind.

## I. INTRODUCTION

In the past most of the attacks targeted a single system, in an attempt to disable it or to gain an unauthorized access to the data in the system. Today, the most dangerous threats no longer attack hosts, but rather the network infrastructure: worms and Distributed Denial of Service (DDOS) attacks launched from bot nets are currently a significant threat, and the protection options remain limited. These attacks are currently used to extort the money from service providers for not launching DOS attacks against their sites.

The main principle of the Internet was openness. Thus, it infrastructure was created under this scheme. However, the price of this success is a weakness in security. In the Internet, anyone can send any packet to anyone without being authenticated, while the receiver has to process any packet that arrives to a provided service. The lack of authentication means that attackers can create a fake identity, and send malicious traffic. All systems connected to the Internet are potential targets for attacks since the openness of the Internet makes them accessible to attack traffic.

One of the most dangerous attacker's activities is a Denial of Service (DoS) attack [1]. DoS attack aims to stop the service provided by a target. It can be launched in two forms. The first form is to exploit software vulnerabilities of a target by sending malformed packets and crash the system. The second form is to use massive volumes of useless traffic to occupy all the resources that could service legitimate traffic. While it is possible to protect the first form of attack by patching known vulnerabilities, the second form of attack cannot be so easily prevented. When the traffic of a DoS attack comes from multiple sources, it called a Distributed Denial of Service (DDoS) attack. By using multiple attack sources, the power of a DDoS attack is amplified and the problem of defense is made more complicated.

## II. OVERVIEW OF ATTACK TYPES

Currently we have numerous DoS attack types. Each attack uses some special exploit of Internet protocols or software weaknesses. For example, these attacks could be launched directly overwhelming by large packets (UDP, ICMP flood), using reflectors (Smurf, Fraggle), and sending too long packets (Ping of Death), wrong packets (Land). Recently one can see progress in this field – new attack types cause damage to computer systems. Novel type of attack, with a low average rate, exploits the transients of a system's dynamic behavior. The low-rate attacks introduce significant inefficiencies that tremendously reduce system capacity or service quality. In the literature, this kind of network assault is called shrew attack [2] or Reduction of Quality (RoQ) attack [3].

Various attacks has special characteristic. In work [4] we propose following parameter set. (Fig 1.):

- Attack type. Security experts segregate DoS attacks into two categories: distributed (attack from many sources) and non-distributed (attack from one computer).
- Attack direction. Attack directions divided on network resources and target resources.
- Attack scheme. Generally speaking, attack scheme can be direct (direct sending of malicious traffic), reflector (traffic reflects from other computers) or hidden (malicious traffic hidden in legal).
- Attack method. Method defines vulnerability that used during attack. Targeted attack uses vulnerability of software, services, and protocols. The primary goal of a consumption attack is to consume all possible available resources (usually network resources) to shut down a system. Exploitive attacks target bugs in operating systems.

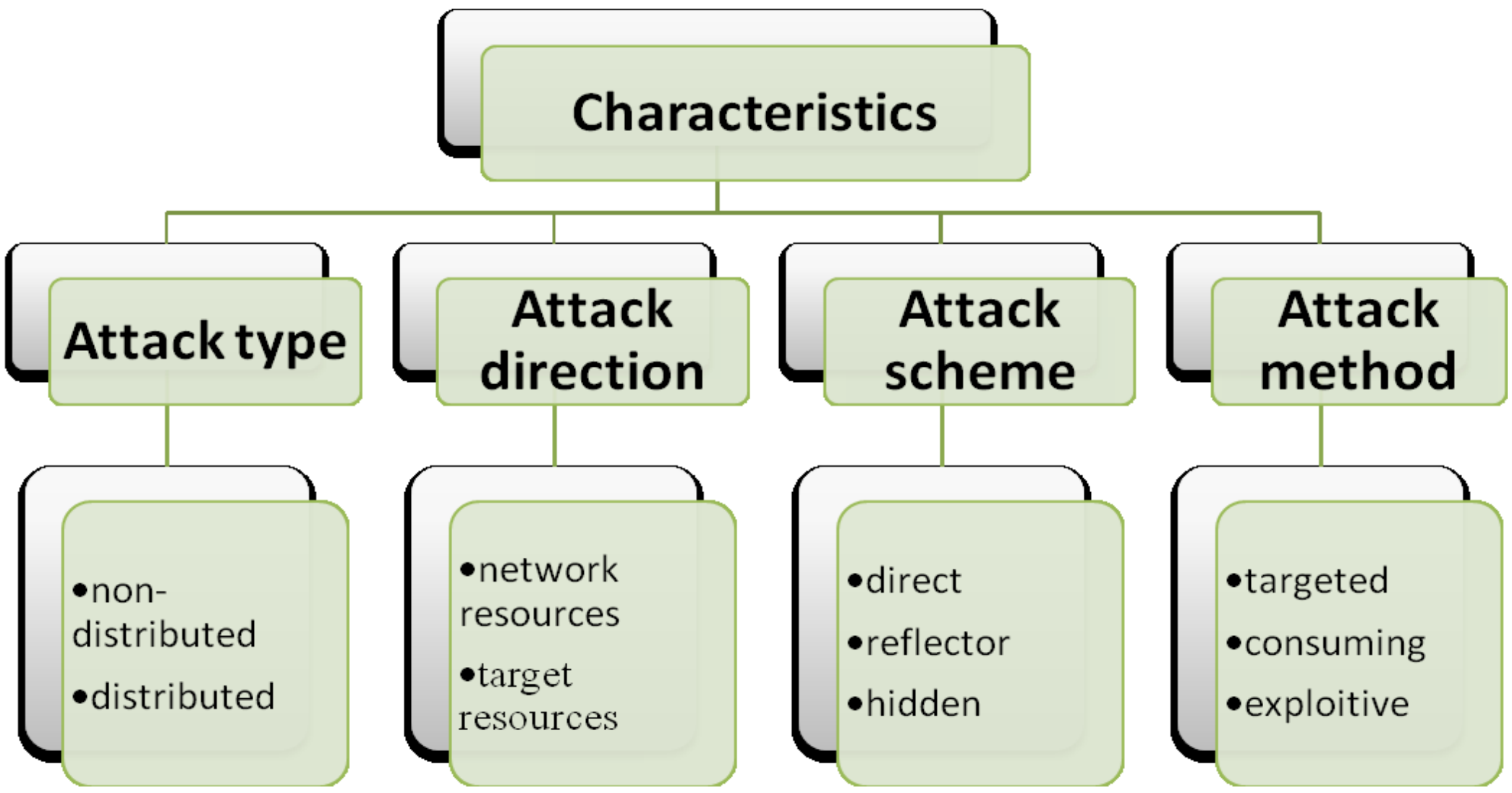

Fig. 1 Attack characteristics.

## III. SYSTEM MODEL

When we build Detection and Reaction System against denial of service attacks one can face with two main problems:

- New attack types. New attacks constantly arise. Moreover they became more intellectual and sophisticated.
- Change in characteristics of defended system. While we build defense system we use some assumption about routine working. If system behavior changes then correct working of a scheme of defense is questionable.

These problems connected with modeling defended system as dynamic control system. When we have such model one can find out if system behavior became anomaly.

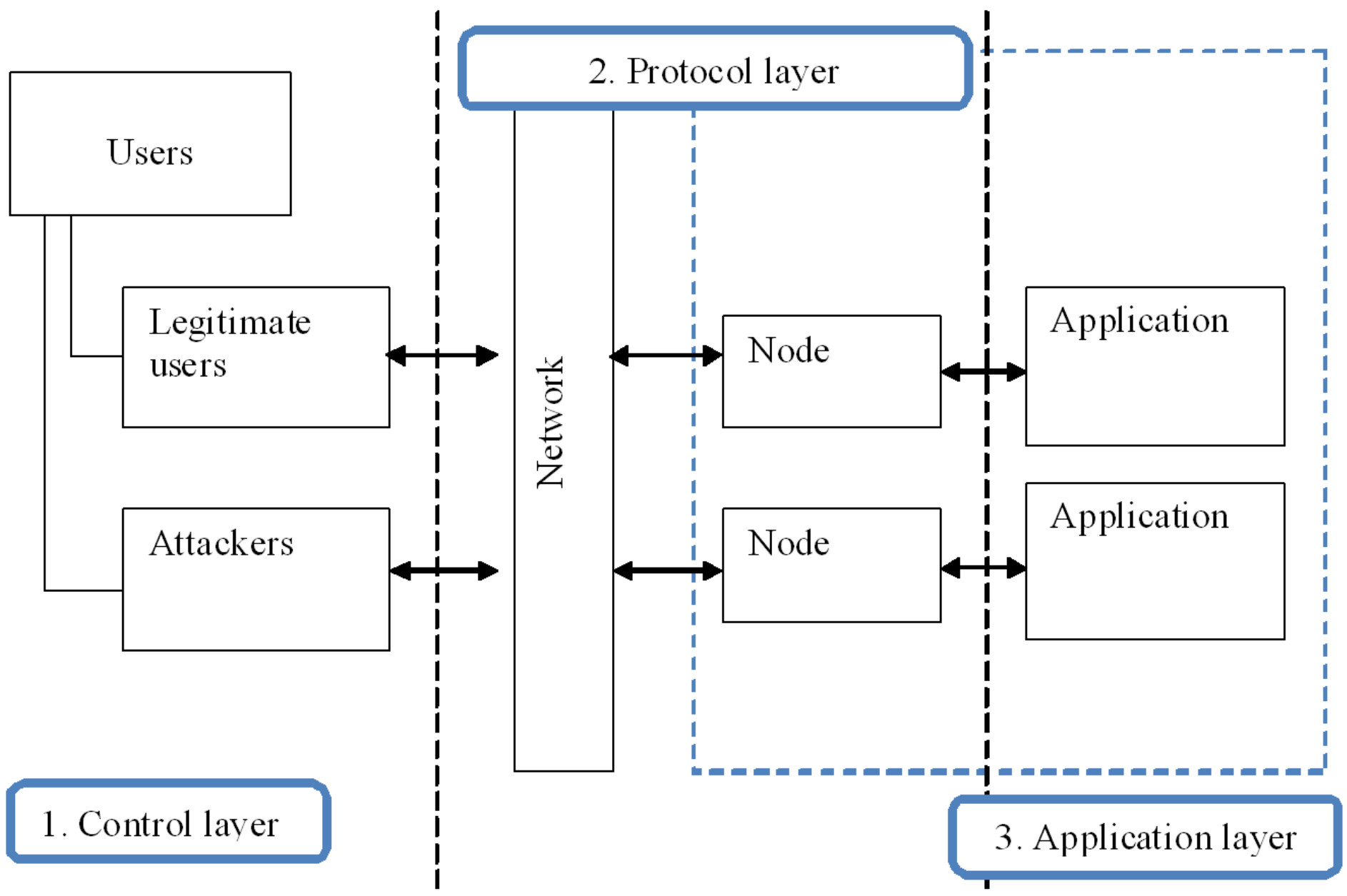

Fig. 2 Three-layer system model.

Let we have computer system connected to the Internet (Fig. 2). System could be divided on three large parts

1. Control layer
2. Protocol layer
3. Application layer

Control layer describe area of users and their actions. There are two categories of users – legitimate users and attackers. Users can change system dynamic by sending control sequence of packets. System cannot forecast user's behavior but can use some trust model.

Protocol layer describe low-level protocol of network. It includes network and nodes. Main characteristics of system dynamic in this part are packet frequency, density, types and so on.

Application layer is responsible for services working. When memory or process time recourses are depleted it is strong indication of anomaly behavior.

Intrusion detection system must gather information from all parts of system and transform it into couple characteristic of anomaly behavior. Currently, intrusion detection systems can be split into two distinct categories. Most commercial and existing open-source products fall into the signature-detection category: these systems match the packets in the network with the predefined set of rules. While this system can be valuable in detecting known attack patterns, it provides no protection against novel threats, or even new permutations of known attack, that are not represented in its knowledge base.

Furthermore, as the rules are typically sparse to achieve efficiency, the signature-based IDS have a non-negligible false positives ratio. Another major disadvantage of the existing systems is a high computational cost of reasonably complete set of rules. The anomaly detection approaches in NIDS (Network IDS) are typically based on classification/aggregation of sessions or flows (unidirectional components of sessions) into classes and deciding whether a particular class contains malicious or legitimate flows. While these approaches are able to detect novel attacks, they suffer from comparably high false-positive (or false negative depending on the model tuning) rate. Most research in the NIDS area currently delivers a combination of signature and anomaly detection and aims to improve the effectiveness and efficiency of intrusion detection.

Let us define typical architecture of multi-agent Intrusion Detection system [5]. The goal of the proposed architecture is to provide a near-real time autonomous attack discovery and response, replacing the direct human decision-making by well defined policies regulating autonomous runtime system decisions. The architecture is distributed, based on a peer-to-peer cooperation of autonomous agents:

• Detection and Reaction Agents (DRA) is the intelligent core of the system. They are located on appropriate hosts or network elements in the protected network and are responsible for collaborative traffic analysis, attack detection and planning of the reaction. They incorporate the trust model (or similar technology) used for attack detection and classification. Their decision-making in the reaction phase is governed by user-defined policies.

• Network Sensors (NS) are specialized agents that observe the traffic on the network, perform the low level analysis and feature extraction and inform the DRA's about their observations. They may also directly detect known malicious traffic (by means of signature detection) and raise an alarm. If appropriate, they may be collocated with DRA's.

• Host Sensors (HS) are the agents that reside on the host and are able to raise an alarm when they suspect an intrusion attempt. This alarm is then used for evaluation of the past traffic towards the host by the DRA's.

• Reconfigurable Network Elements (NE) are used by DRA agents to implement the protection mechanism when a new kind of attack is detected.

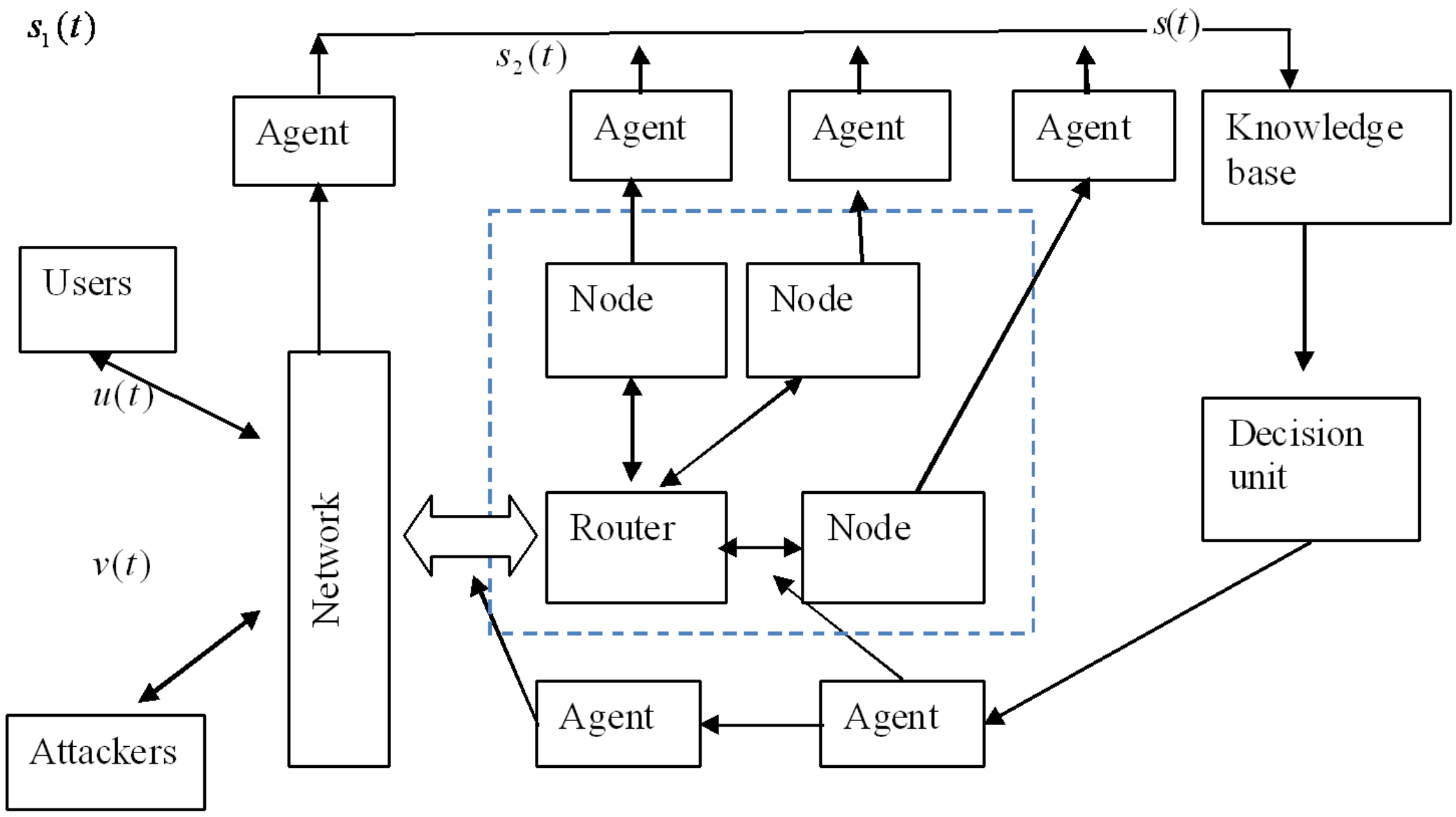

Fig. 3 Agent Detection and Reaction System.

The agents in the community operate in three conceptual phases (Fig. 3). In the first phase, the NS agents observe the traffic on the network and report the features of the relevant observed network flows to the DRA. The DRA agents then use their trust model to classify unknown traffic by matching it with a reaction from the Host Sensors. To perform the match in the second phase of the detection, the existing trust models must be considerably extended. Trustfulness values are therefore no longer relative to the real identities or individual connections, but are attached to the centers of the clusters created in the metric space of the features.

When a new connection appears, we use a clustering algorithm to either attach it to an existing cluster, or to create a new cluster altogether. To update the trustfulness of the individual clusters, we rely on a feedback of the HS agents. Once aggregated, the aggregated feedback from related HS is used as a new observation of connection trustfulness for all open and recent relevant connections. One of the important assumptions we make is that the protected system contains many heterogeneous hosts with a variety of Host Sensors. Therefore, given the random propagation strategy of typical worm attacks, we shall be able to identify the attack on the hosts that are not vulnerable and are appropriately protected.

Each DRA maintains its own instance of the trust model, and can extend the basic shared feature space with unique features of its own. To communicate these values between the DRA's, the agents use a reputation mechanism in both subscribe - inform and query mode. Such mechanism will ensure that once the attack is suspected by a single DRA, its classification will spread through the system and allow the other DRA to react faster. On the other hand, they are still autonomous to react differently, reducing the risk of false positives. In the third phase, occurring after the attack detection, the DRA's need to create its description from the model, typically using generalizing machine learning methods.

## REFERENCES

[1] Igor Kotenko. Multi-agent Simulation of Attacks and Defense Mechanisms in Computer Networks. The Journal of Computing, Vol. 7, Issue 2, 2008. P.35-43.

[2] Y.-K. Kwok, R. Tripathi, Y. Chen, and K. Hwang, "HAWK: Halting Anomaly with Weighted ChoKing to Rescue Well-Behaved TCP Sessions from Shrew DoS Attacks," accepted to appear in the 2005 International Conference on Computer Networks and Mobile Computing (ICCNMC 2005), Zhangjiajie, China, August 2-4, 2005.

[3] M. Guirguis, A. Bestavros, I. Matta, and Y. Zhang, "Reduction of Quality (RoQ) Attacks on Internet End Systems," Proc. INFOCOM 2005.

[4] O. Ignatenko, P. Andon. Counteraction to denial of service attacks in Internet: approach concept // Problems of programming, 2-3, 2008, P. 564 – 574.

[5] M. Reh´ak and other. Agent Methods for Network Intrusion Detection and Response. HoloMAS 2007, LNAI 4659, pp. 149–160, 2007.

## IV. CONCLUSION

In this paper mitigation of denial of service attacks is considered. Distributed agent-based intrusion detection system is proposed.